\documentclass[%
 reprint,
 amsmath,amssymb,
 aps,
]{revtex4-2}

\usepackage{graphicx}
\usepackage{dcolumn}
\usepackage{bm}

\begin{document}

\preprint{APS/123-QED}

\title{A Pfaffian formulation for matrix elements of three-body operators in multiple quasi-particle configurations}

\author{Zi-Rui Chen}%
\affiliation{School of Physical Science and Technology, Southwest University, Chongqing 400715, China}%

\author{Long-Jun Wang}
\email{longjun@swu.edu.cn}
\affiliation{School of Physical Science and Technology, Southwest University, Chongqing 400715, China} 

\date{\today}

\begin{abstract}
  We present a Pfaffian formula to calculate matrix elements of three-body operators in symmetry-restoration beyond-mean-field methods, including the case of multiple quasi-particle configurations. Detailed derivation based on [Mizusaki \emph{et al.}, Phys. Lett. B 715, 219 (2012)] and [Hu \emph{et al.}, Phys. Lett. B 734, 162 (2014)] is provided, and potential applications in generator coordinate method with chiral interactions, as well as in study of nuclear matrix elements in neutrinoless double beta decay are discussed. 
\end{abstract}

\maketitle


\section{\label{sec:intro}Introduction}

Successful descriptions of nuclear structures rely heavily on the solution of nuclear many-body problem \cite{Ring_many_body_book}, which is difficult due to two aspects: the complexity of nucleon-nucleon interaction and the troublesome many-body techniques. On one hand, fundamental progress has been achieved during the past decades for modern understandings of nuclear forces, based on, for example, the chiral effective field theory \cite{nuclear_force_2009_RMP, chiral_EFT_2011_Phys_Rep} where three-body nuclear forces are found to play important roles in nuclear structure physics.

A lot of nuclear many-body techniques, on the other hand, employ the philosophy that reduces the nuclear many-body to effective one-body problems with the help of the concept of quasiparticles (qp) and single-particle mean-field calculations in the intrinsic systems, such as the Hartree-Fock-Bogoliubov (HFB) theory \cite{Heyde_Basic_idea_book, Suhonen_book, Bender_2003_RMP}. Residual nucleon-nucleon correlations could be included reasonably through beyond-mean-field methods, which provide description of nuclear many-body wave functions in the laboratory frame. 

One kind of the popular nuclear models, the generator coordinate method (GCM) \cite{Ring_many_body_book, GCM_Yao_2010, GCM_IMSRG_Yao_2018, LJWang_current_2018_Rapid, GCM_Bender_2008, GCM_Tomas, GCM_Chen_2013, GCM_Nobuo_2014, GCM_Hagino_2021, GCM_Shimizu_2021, GCM_qp_FQChen_2017, GCM_Higashiyama_2021, GCM_Jiao_2019, GCM_Robledo_2021} or related angular-momentum projection (AMP) based method \cite{PSM_review, PSM_Sun2, TPSM_1999_PRL, Zhao_P_W_projection_2016, TPSM_Sheikh_2011_PRC, Calvin_Johnson_2017_PRC, Calvin_Johnson_2018_JPG, Calvin_Johnson_2021_JPG, Jia_Qi_2018_PRC, Sheikh_2021_JPG_review}, usually starts from single-particle HFB calculations in the intrinsic system where some symmetries are broken, which can be restored exactly by the projection technique from which the description of nuclear systems in the laboratory frame can be achieved. Further nucleon-nucleon correlations are then included by diagonalizing Hamiltonian in the non-orthonormal projected basis, which leads to the solution of the Hill-Wheeler-Griffin equation. The central ingredients (kernels) of these models turn out to be different projected (or \emph{rotated}) matrix elements generated by AMP, particle-number projection (PNP), parity projection etc. where the AMP usually dominates analytical and numerical efforts. During the past decades, the GCM or AMP (PNP) based methods have been applied successfully to researches on nuclear low-lying states \cite{GCM_Yao_2010, GCM_Bender_2008, GCM_Tomas, GCM_Chen_2013, Fu_2018_PRC}, high-spin physics \cite{PSM_review, TPSM_Sheikh_2011_PRC, Zhao_P_W_projection_2016, Calvin_Johnson_2017_PRC, LJWang_2014_PRC_Rapid, LJWang_2016_PRC, LJWang_PLB_2020_chaos, Yokoyama_2021_PRC_Lett}, $\beta$-decay \cite{LJWang_2018_PRC_GT}, neutrinoless double-$\beta$ decay \cite{GCM_Nobuo_2014, GCM_Jiao_2019, LJWang_current_2018_Rapid, Tomas_2010_PRL_0vbb, Yao_2015_PRC_0vbb, Y_K_Wang_PRC_2021_0vbb}, astrophysical weak process \cite{LJWang_PLB_2020_ec, LJWang_2021_PRL}, nuclear fission \cite{Fission_Bertsch_Robledo_2019_PRC, Robledo_2018_JPG_review, TD_GCM_2016_cpc, TD_GCM_2018_cpc, TD_GCM_2019_PRC} etc. with different effective interactions or schematic interactions. 

In the above applications, collective degrees of freedom (such as shape fluctuations) \cite{GCM_Yao_2010, GCM_Bender_2008, GCM_Tomas}, single-particle (such as qp excitations) degrees of freedom \cite{PSM_review, Zhao_P_W_projection_2016, LJWang_2014_PRC_Rapid} or both of them \cite{GCM_qp_FQChen_2017} are included, according to the underlying physical problems of interests and the practical computational burden simultaneously. The computational burden concentrates on the kernels in the Hill-Wheeler-Griffin equation, i.e., the \emph{rotated} norm overlap of HFB qp vacuum, the \emph{rotated} norm overlap of multi-qp configuration, and the \emph{rotated} matrix elements of Hamiltonian (including one-body, two-body and potentially three-body operators). Historically, these three kinds of kernels could be calculated by the Onishi formula \cite{Onishi}, the generalized Wick's theorem \cite{PSM_review}, Hara's prescription (see the Appendix of Ref. \cite{PSM_review}) etc. which, unfortunately, encountered the sign problem, the problem of combinatorial complexity and extremely computational cost, respectively. 

During the past decade or so, after the pioneering work of Robledo \cite{Robledo_2009_PRC} who solved the sign problem of the Onishi formula in a mathematically elegant way in terms of Pfaffian by making use of Grassmann numbers and Fermion coherent state, the Pfaffian formulations (algorithms) have been developed rapidly for all the three kinds of kernels in the Hill-Wheeler-Griffin equation \cite{Robledo_2009_PRC, Robledo_2011_PRC, Pfaffian_code_CPC_2011, Bertsch_2012_PRL, Oi_2012_PLB, Mizusaki_2012_PLB, Avez_2012_PRC, Mizusaki_2013_PLB, Gao_2014_PLB, Hu_2014_PLB, Mizusaki_2018_PLB, Carlsson_2021_PRL}. In particular, the problem of combinatorial complexity for the norm overlap of multi-qp configuration can be avoided with the help of the Pfaffian formula by Mizusaki \emph{et al.} \cite{Mizusaki_2013_PLB} and the calculation of \emph{rotated} matrix elements of one-body and two-body operators can be optimized to a large extent according to the Pfaffian formula by Hu \emph{et al.} \cite{Hu_2014_PLB}. These achievements of Pfaffian formulations, then, would make GCM method one of the optimal nuclear many-body techniques from both physical (with broken symmetries restored) and numerically practical perspectives.

Recently, the GCM methods based on realistic shell-model Hamiltonians have been developed for studies on nuclear low-lying states and neutrinoless double-$\beta$ decay \cite{CFJiao_2017_PRC, CFJiao_2018_PRC, Calvin_Johnson_2021_JPG, GCM_Shimizu_2021}, and the in-medium similarity-renormalization group (IMSRG) method has been updated by employing symmetry-restored states or GCM calculations as the reference states with shell-model Hamiltonian \cite{GCM_IMSRG_Yao_2018} and the chiral Hamiltonian \cite{Yao_2020_PRL} to describe low-lying states of deformed light nuclei and the neutrinoless double-$\beta$ decay among them. On the other hand, for many interesting nuclear-structure and decay problems such as the high-spin physics and astrophysical weak-interaction process, \emph{ab initio} methods for both low-lying and relatively highly excited states of medium-heavy and heavy nuclei are demanded. One of the potential candidates may be a GCM approach with both collective (such as shape fluctuations) and single-particle (such as multi-qp configurations) degrees of freedom by realistic nuclear forces such as the chiral forces. The evaluation of matrix elements of three-body operators is then indispensable. Besides, for studies on the uncertainty from transition (decay) operator in nuclear matrix elements for neutrinoless double-$\beta$ decays, chiral two-body currents needs to be considered and matrix elements of three-body transition (decay) operator \cite{LJWang_current_2018_Rapid} in beyond-mean-field approached are also demanded. 

In this work, we provide a Pfaffian formula for evaluation of matrix elements of general three-body operators in symmetry-restoration beyond-mean-field methods such the GCM or AMP-based models, for cases with or without multiple qp configurations. In Sec. \ref{sec:theory} we briefly introduce the basic logics of GCM or AMP (PNP) based models. In Sec. \ref{sec:result} we provide the Pfaffian formula for evaluation of matrix elements of three-body operators, and we finally summarize our work in Sec. \ref{sec:sum}.

\section{\label{sec:theory}Basic logic of GCM or AMP-based models}

For completeness of discussion, we first give a concise introduction of the logic of GCM or AMP(PNP)-based models \cite{Ring_many_body_book}. The starting point of GCM models is a description of nuclear many-body systems in the intrinsic system by solving the single-nucleon HFB mean-field equation with constraints (on quantities with respect to related coordinates $\bm q$, such as the total quadrupole moments etc.), from which one can get roughly approximate many-body wave functions for the ground state and (qp) excited states as,
\begin{eqnarray} \label{config}
  |\Phi_\kappa(\bm q) \rangle = \{ |\Phi(\bm q)\rangle, \ \hat\beta^\dag_i(\bm q) \hat\beta^\dag_j(\bm q) |\Phi(\bm q)\rangle, \ \cdots \}
\end{eqnarray}
where $|\Phi(\bm q)$ labels the HFB qp vacuum and the corresponding qp operators are denoted by $\{\hat\beta_i(\bm q), \hat\beta^\dag_i(\bm q)\}$. In Eq. (\ref{config}) the index $\kappa$ reflects the information of qp excitations. 

The wave functions in Eq. (\ref{config}) broke some symmetries which can be recovered by projection operators. Here we take the AMP and PNP operators as examples since the parity projections etc. are relatively trivial. The PNP operator is,
\begin{eqnarray} \label{pnp}
  \hat P^\tau = \frac{1}{2\pi} \int_0^{2\pi} d\phi_\tau e^{i(\hat N_\tau - N_\tau) \phi_\tau } ,
\end{eqnarray}
where $\hat N_\tau$ is the particle-number operator for neutrons ($\tau=n$) or protons ($\tau=p$) and $\phi_\tau$ is the gauge angle. The AMP operator reads as,
\begin{eqnarray} \label{amp}
  \hat P^J_{MK} = \frac{2J+1}{8\pi^2} \int d\Omega D^{J\ast}_{MK}(\Omega) \hat R(\Omega),
\end{eqnarray}
with $D^J_{MK}$ being the Wigner $D$ function, $\hat R$ the rotation operator with respect to the Euler angle $\Omega$.

The AMP and PNP operators can recover the broke symmetries in $|\Phi_\kappa(\bm q) \rangle$ and provide description of nuclear many-body systems in the laboratory frame, i.e., 
\begin{eqnarray} \label{project_basis}
  \hat P^{J,NZ}_{MK} |\Phi_\kappa(\bm q) \rangle \equiv \hat P^J_{MK} \hat P^N \hat P^Z |\Phi_\kappa(\bm q) \rangle,
\end{eqnarray}
Besides, more nucleon-nucleon correlations could be included by diagonalizing Hamiltonian in the non-orthonormal projected basis in Eq. (\ref{project_basis}), so that one can write the nuclear many-body wave functions as,
\begin{eqnarray} \label{final_wave_func}
  |\Psi^\sigma_{JM} \rangle = \int d\bm q \sum_{K\kappa} f^{J \sigma}_{K\kappa}(\bm q) \hat P^{J,NZ}_{MK} |\Phi_\kappa(\bm q) \rangle,
\end{eqnarray}
where $\sigma$ denotes the $\sigma^{\text{th}}$ eigen state for angular momentum $J$. This corresponds to the solution of the Hill-Wheeler-Griffin equation,
\begin{eqnarray} \label{Hill_Wheeler}
  \sum_{K'\kappa' \bm q'} \left[ \mathcal H^{J}_{K\kappa K'\kappa'}(\bm q, \bm q') - E^{\sigma}_J \mathcal N^{J}_{K\kappa K'\kappa'}(\bm q, \bm q') \right] f^{J \sigma}_{K'\kappa'}(\bm q') = 0, \nonumber \\
\end{eqnarray}
from which the coefficient $f^{J \sigma}_{K\kappa}(\bm q)$ in Eq. (\ref{final_wave_func}) can be obtained and the nuclear many-body wave functions $|\Psi^\sigma_{JM} \rangle$ can be well defined. 

Then, physical quantities in the laboratory frame can be calculated and compared with measurements. Let us take typical transitions and decays as examples. With the corresponding operator $\mathcal T^{\lambda \mu}$, the (reduced) transition strengths could be obtained by means of,
\begin{eqnarray} \label{T_operator}
  \left\langle \Psi^{(\mathcal S)\sigma}_{JM} \Big| \mathcal T^{\lambda \mu} \Big| \Psi^{(\mathcal S')\sigma'}_{J'M'} \right\rangle ,
\end{eqnarray}
where $(\mathcal S)$ and $(\mathcal S')$ can represent the same nuclear system (such as, for electromagnetic transitions) or two different nuclear systems (such as, for $\beta$ decay and double $\beta$ decay etc.).

In Eq. (\ref{Hill_Wheeler}) $\mathcal H$ and $\mathcal N$ read as,
\begin{subequations} \label{H_and_N}
\begin{eqnarray}
  \mathcal H^{J}_{K\kappa K'\kappa'}(\bm q, \bm q') & = & \langle \Phi_\kappa(\bm q) | \hat H \hat P^{J,NZ}_{KK'} | \Phi_{\kappa'}(\bm q') \rangle,  \\
  \mathcal N^{J}_{K\kappa K'\kappa'}(\bm q, \bm q') & = & \langle \Phi_\kappa(\bm q) |  P^{J,NZ}_{KK'} | \Phi_{\kappa'}(\bm q') \rangle .
\end{eqnarray}
\end{subequations}

As mentioned in the Introduction, the central ingredients (kernels) of the GCM models are three kinds of \emph{rotated} matrix elements generated by projection techniques. From Eqs. (\ref{pnp}, \ref{amp}, \ref{final_wave_func}, \ref{T_operator}, \ref{H_and_N}a, \ref{H_and_N}b) it is seen that the three kinds of (kernels) \emph{rotated} matrix elements are,
\begin{subequations} \label{kernels} 
\begin{eqnarray}
  \langle \Phi^a | & \hat{\mathcal R} & | \Phi^b \rangle, \\
  \langle \Phi^a | \hat \alpha_1 \cdots \hat \alpha_m & \hat{\mathcal R} & \hat \beta^\dag_1 \cdots \hat \beta^\dag_{m'} | \Phi^b \rangle, \\
  \langle \Phi^a | \hat \alpha_1 \cdots \hat \alpha_m & \hat{\mathcal O} \hat{\mathcal R} & \hat \beta^\dag_1 \cdots \hat \beta^\dag_{m'} | \Phi^b \rangle.
\end{eqnarray}
\end{subequations}
i.e., the norm overlap of HFB qp vacuum, the norm overlap of multi-qp configuration, and the \emph{rotated} matrix elements of operators, respectively. In Eq. (\ref{kernels}), the indices $a \equiv \{\mathcal S, \bm q \}$ and $b \equiv \{\mathcal S', \bm q' \}$ can represent the same or different nuclear systems with corresponding qp operators $\hat\alpha_i$ and $\hat\beta_j$, and $\hat{\mathcal R}$ labels the total unitary (\emph{rotation}) operator, i.e., $\hat{\mathcal R} \equiv \hat R(\Omega) e^{i \hat N_\tau }$ in the AMP+PNP case considered here. The operator $\hat{\mathcal O}$ can be the Hamiltonian (one-body, two-body and even three-body scalar operators) and the transition or decay operators (one-body, two-body and even three-body tensor operators).

Fortunately, the norm overlap of HFB qp vacuum in Eq. (\ref{kernels}a) can be calculated by the Pfaffian formula in Refs. \cite{Robledo_2009_PRC, Robledo_2011_PRC, Bertsch_2012_PRL, Oi_2012_PLB, Avez_2012_PRC, Gao_2014_PLB, Mizusaki_2018_PLB, Carlsson_2021_PRL} avoiding the notorious sign problem in the Onishi formula, the norm overlap among multi-qp configurations in Eq. (\ref{kernels}b) can be evaluated with the help of the Pfaffian formula in Refs. \cite{Bertsch_2012_PRL, Mizusaki_2012_PLB, Mizusaki_2013_PLB, Hu_2014_PLB} which do not suffer from the problem of combinatorial complexity in the generalized Wick's theorem any more, and the Pfaffian formula for evaluation of \emph{rotated} matrix elements of one-body and two-body operators are provided in Ref. \cite{Hu_2014_PLB}. In the next section, we follows the techniques in Refs. \cite{Mizusaki_2012_PLB, Hu_2014_PLB} to provide the Pfaffian formula for \emph{rotated} matrix elements of three-body operators.

\section{\label{sec:result}The Pfaffian formulation for three-body operators}

As mentioned previously, the Pfaffian formula for \emph{rotated} matrix elements of one-body and two-body operators are derived by Hu \emph{et al.} \cite{Hu_2014_PLB}. The derivations are achieved in a much mathematical way by adopting the expansion properties of the Pfaffian with respect to rows and columns. In the following we adopt the similar techniques to derive the Pfaffian formula for \emph{rotated} matrix elements of three-body operators, and discuss the underlying physics and treatments in potential applications in nuclear structure physics. In the second quantization representation we can write three-body operators as,
\begin{eqnarray} \label{3b_operator}
  \hat{V}^{(3)} = \sum_{\mu\nu\delta \omega\rho\gamma}^M W_{\mu\nu\delta \omega\rho\gamma} \ 
  \hat{c}^\dag_{\mu} \hat{c}^\dag_{\nu} \hat{c}^\dag_{\delta} \ \hat{c}_{\gamma} \hat{c}_{\rho} \hat{c}_{\omega},
\end{eqnarray}
where the factor $1/36$ is absorbed to the anti-symmetric matrix $W$, $\hat c^\dag$ and $\hat c$ denote the particle creation and annihilation operators in the spherical harmonic oscillator basis and $M$ labels the dimension of the single-particle model space.

To derive and better understand the Pfaffian formula for physical operator, we need to rely on the following generalized Pfaffian formula of norm overlaps which is derived in Ref. \cite{Hu_2014_PLB} and which actually corresponds to the generalized Wick's theorem, i.e., 
\begin{eqnarray} \label{Pf_norm}
  \langle \Phi | \hat z_1 \cdots \hat z_{2N} | \Phi' \rangle = \text{Pf}(\mathbb S) \langle \Phi |\Phi' \rangle,
\end{eqnarray} 
where $\mathbb S$ is a $2N\times 2N$ skew-symmetric matrix with the elements
\begin{eqnarray} \label{the_S_ij}
  \mathbb S_{ij} \equiv \frac{\langle \Phi | \hat z_i \hat z_j | \Phi' \rangle}{\langle \Phi | \Phi' \rangle} \qquad (i<j).
\end{eqnarray} 
Note that in Eq. (\ref{Pf_norm}) $ \langle \Phi |\Phi' \rangle \ne 0$ is assumed \cite{Hu_2014_PLB} and can be calculated by the Pfaffian formulae in the literatures \cite{Robledo_2009_PRC, Robledo_2011_PRC, Bertsch_2012_PRL, Oi_2012_PLB, Avez_2012_PRC, Gao_2014_PLB, Mizusaki_2018_PLB, Carlsson_2021_PRL}. $| \Phi \rangle$ and $| \Phi' \rangle$ can be the same or different HFB vacua (as in Eqs. (\ref{config}, \ref{kernels})), or some unitary transformation of HFB vacua (see below), or even the true vacuum $|-\rangle$. Besides, the single-fermion operators $\hat z_i$ could be the qp creation (annihilation) operators for either $| \Phi \rangle$ or $| \Phi' \rangle$, or any unitary transformation between them (such as the particle operators $\hat c^\dag$ and $\hat c$ etc.). Eq. (\ref{Pf_norm}) is equivalent to the generalized Wick's theorem that considers all possible contractions among $\hat z_i$ (with $(2N-1)!!$ contractions in total). The norm overlap of multi-qp configuration in Eq. (\ref{kernels}b) represents a simple example of Eq. (\ref{Pf_norm}). 

As illustrated by Eqs. (\ref{Hill_Wheeler}, \ref{T_operator}, \ref{H_and_N}, \ref{kernels}) we now treat the \emph{rotated} matrix elements of three-body operators
\begin{eqnarray} \label{3b_ME}
I_3 &=& \sum_{\mu\nu\delta \omega\rho\gamma}^M W_{\mu\nu\delta \omega\rho\gamma} \times \nonumber \\
    & & \langle \Phi^a |   \hat \alpha_1 \cdots \hat \alpha_L   
        \hat{c}^\dag_{\mu} \hat{c}^\dag_{\nu} \hat{c}^\dag_{\delta} \ \hat{c}_{\gamma} \hat{c}_{\rho} \hat{c}_{\omega}
        \hat{\mathcal R}  \hat \beta^\dag_{L+1} \cdots \hat \beta^\dag_{2N}
        | \Phi^b \rangle 
\end{eqnarray}
which can be written as
\begin{eqnarray} \label{3b_ME_b}
I_3 &=& \sum_{\mu\nu\delta \omega\rho\gamma}^M W_{\mu\nu\delta \omega\rho\gamma} \times \nonumber \\
    & & \langle \Phi |   \hat z_1 \cdots \hat z_L   
        \hat{c}^\dag_{\mu} \hat{c}^\dag_{\nu} \hat{c}^\dag_{\delta} \ \hat{c}_{\gamma} \hat{c}_{\rho} \hat{c}_{\omega}
        \hat z_{L+1} \cdots \hat z_{2N}
        | \Phi' \rangle
\end{eqnarray}
by defining 
\begin{subequations}
\begin{eqnarray}
  \hat z_k &=& \left\{ \begin{array}{ll}
               \hat \alpha_k, & 1 \leqslant k \leqslant L \\
               \hat{\mathcal R}  \hat \beta^\dag_{k} \hat{\mathcal R}^{-1}, & L+1 \leqslant k \leqslant 2N
               \end{array} \right. \\
  | \Phi  \rangle &=& | \Phi^a \rangle, \\
  | \Phi' \rangle &=& \hat{\mathcal R} | \Phi^b \rangle.
\end{eqnarray}
\end{subequations}

Now the evaluation of \emph{rotated} matrix elements for physical operators is much straightforward actually. Taking the three-body operator case $I_3$ as an example, one can calculate each $\langle \Phi | \hat z_1 \cdots \hat z_L   \hat{c}^\dag_{\mu} \hat{c}^\dag_{\nu} \hat{c}^\dag_{\delta}  \hat{c}_{\gamma} \hat{c}_{\rho} \hat{c}_{\omega} \hat z_{L+1} \cdots \hat z_{2N}  | \Phi' \rangle$, multiplied by corresponding $W_{\mu\nu\delta \omega\rho\gamma}$ and finally consider the 6-fold loops (summations) for indices $\{\mu\nu\delta \omega\rho\gamma\}$ in Eq. (\ref{3b_ME_b}). The $\langle \Phi | \hat z_1 \cdots \hat z_L   \hat{c}^\dag_{\mu} \hat{c}^\dag_{\nu} \hat{c}^\dag_{\delta}  \hat{c}_{\gamma} \hat{c}_{\rho} \hat{c}_{\omega} \hat z_{L+1} \cdots \hat z_{2N}  | \Phi' \rangle$ can be calculated by the generalized Wick's theorem, or equivalently by the Pfaffian formula in Eq. (\ref{Pf_norm}) for which a $(2N+6)\times(2N+6)$ skew-symmetric matrix (say $\mathbb M$) should be defined first. In either of the two ways, basic contractions among $\{\hat z_i, \hat c^\dag_\mu, \hat c_\nu \}$ (the matrix elements of $\mathbb M$) should be calculated in advance. 
The basic contractions among $\{\hat z_i\}$ themselves have been defined in Eq. (\ref{the_S_ij}). For the basic contractions involving $\hat c^\dag$ and/or $\hat c$ we define them in the following 
 
\begin{subequations} \label{SSCCC}
\begin{eqnarray}
  \mathbb{S}_{\mu k}^{(+)} 
    &=& \left \{ \begin{array}{cl}
        - \frac{\langle \Phi | \hat{z}_k \hat{c}_{\mu}^{\dag} |\Phi' \rangle}{\langle \Phi| \Phi' \rangle}  & \quad 1 \leqslant k \leqslant L \\
        \frac{ \langle \Phi | \hat{c}_{\mu}^{\dag} \hat{z}_k |\Phi' \rangle}{\langle \Phi |\Phi' \rangle}   & \quad L+1 \leqslant k \leqslant 2N\\
        \end{array}
        \right. , \\
  \mathbb{S}_{\mu k}^{(-)} 
    &=& \left \{ \begin{array}{cl}
        - \frac{\langle \Phi | \hat{z}_k \hat{c}_{\mu} |\Phi' \rangle}{\langle \Phi| \Phi' \rangle}   & \quad 1 \leqslant k \leqslant L \\
        \frac{ \langle \Phi | \hat{c}_{\mu} \hat{z}_k |\Phi' \rangle}{\langle \Phi |\Phi' \rangle}    & \quad L+1 \leqslant k \leqslant 2N\\
        \end{array}
        \right. , \\
  \mathbb{C}^{(+)}_{\mu\nu} &=& \frac{\langle \Phi | \hat{c}^\dag_\mu \hat{c}^\dag_{\nu} |\Phi' \rangle}{\langle \Phi| \Phi' \rangle}, \\
  \mathbb{C}^{(0)}_{\mu\nu} &=& \frac{\langle \Phi | \hat{c}^\dag_\mu \hat{c}_{\nu} |\Phi' \rangle}{\langle \Phi| \Phi' \rangle}, \\
  \mathbb{C}^{(-)}_{\mu\nu} &=& \frac{\langle \Phi | \hat{c}_\mu \hat{c}_{\nu} |\Phi' \rangle}{\langle \Phi| \Phi' \rangle}.
\end{eqnarray}
\end{subequations}

Although one can calculate $I_3$ in the above straightforward way, it should be much time-consuming due to the 6-fold loops in large model space and the non-negligible CPU time for computation of the Pfaffian of $(2N+6)\times(2N+6)$ matrix \cite{Pfaffian_code_CPC_2011} for each loops. In the following we derive a compact form for the evaluation of $I_3$ in terms of Pfaffians. The derivation can be done by either Hara's prescription or the expansion properties of Pfaffian with respect to six neighboring rows. We would adopt the former way and leave the equivalent latter way in the Appendix. 

For the $\langle \Phi | \hat z_1 \cdots \hat z_L   \hat{c}^\dag_{\mu} \hat{c}^\dag_{\nu} \hat{c}^\dag_{\delta} \hat{c}_{\gamma} \hat{c}_{\rho} \hat{c}_{\omega} \hat z_{L+1} \cdots \hat z_{2N}  | \Phi' \rangle$ in the calculations of $I_3$ in Eq. (\ref{3b_ME_b}), by either the generalized Wick's theorem or the Pfaffian formula in Eq. (\ref{Pf_norm}) we need to consider $(2N+6-1)!!$ terms, each of the terms corresponds to a possible contraction way for $\{\hat z, \hat c^\dag, \hat c\}$. From Hara's prescription one can classify these $(2N+6-1)!!$ terms into four classes, and get
\begin{eqnarray} \label{I3_sum} 
  \frac{I_3}{\langle \Phi | \Phi' \rangle} = O^{(0)} + O^{(1)} + O^{(2)} + O^{(3)} ,
\end{eqnarray}
The first class $O^{(0)}$ corresponds to contractions among $\{ \hat{c}^\dag_{\mu} \hat{c}^\dag_{\nu} \hat{c}^\dag_{\delta} \hat{c}_{\gamma} \hat{c}_{\rho} \hat{c}_{\omega} \}$ themselves multiplied by contractions among $\{\hat z_i, \ 1 \leqslant i \leqslant 2N \}$, i.e.,
\begin{eqnarray} \label{O0}
  O^{(0)} = W_{0} \text{Pf}( \mathbb{S} ),
\end{eqnarray}
where
\begin{eqnarray}
  W_{0} = \sum_{\mu \nu\delta\omega\rho\gamma} W_{\mu\nu\delta\omega\rho\gamma} \mathbb{C}_{\mu \nu\delta\gamma\rho\omega },
\end{eqnarray}
\begin{eqnarray}
  \mathbb{C}_{\mu \nu\delta \gamma\rho\omega }
  &=& \mathbb{C}^{(+)}_{\mu \nu } \mathbb{C}^{(0)}_{\delta \gamma }\mathbb{C}^{(-)}_{\rho \omega } - 
      \mathbb{C}^{(+)}_{\mu \nu }\mathbb{C}^{(0)}_{\delta \rho }\mathbb{C}^{(-)}_{\gamma \omega } +
      \mathbb{C}^{(+)}_{\mu \nu }\mathbb{C}^{(0)}_{\delta \omega }\mathbb{C}^{(-)}_{\gamma \rho } \nonumber\\
  &-& \mathbb{C}^{(+)}_{\mu \delta }\mathbb{C}^{(0)}_{\nu \gamma }\mathbb{C}^{(-)}_{\rho \omega } +
      \mathbb{C}^{(+)}_{\mu \delta }\mathbb{C}^{(0)}_{\nu \rho }\mathbb{C}^{(-)}_{\gamma \omega } -
      \mathbb{C}^{(+)}_{\mu \delta }\mathbb{C}^{(0)}_{\nu \omega }\mathbb{C}^{(-)}_{\gamma \rho } \nonumber\\
  &+& \mathbb{C}^{(0)}_{\mu \gamma }\mathbb{C}^{(+)}_{\nu \delta }\mathbb{C}^{(-)}_{\rho \omega } - 
      \mathbb{C}^{(0)}_{\mu \gamma }\mathbb{C}^{(0)}_{\nu \rho }\mathbb{C}^{(0)}_{\delta \omega } +
      \mathbb{C}^{(0)}_{\mu \gamma }\mathbb{C}^{(0)}_{\nu \omega }\mathbb{C}^{(0)}_{\delta \rho } \nonumber\\
  &-& \mathbb{C}^{(0)}_{\mu \rho }\mathbb{C}^{(+)}_{\nu \delta }\mathbb{C}^{(-)}_{\gamma \omega } +
      \mathbb{C}^{(0)}_{\mu \rho }\mathbb{C}^{(0)}_{\nu \gamma }\mathbb{C}^{(0)}_{\delta \omega } -
      \mathbb{C}^{(0)}_{\mu \rho }\mathbb{C}^{(0)}_{\nu \omega }\mathbb{C}^{(0)}_{\delta \gamma } \nonumber\\
  &+& \mathbb{C}^{(0)}_{\mu \omega }\mathbb{C}^{(+)}_{\nu \delta }\mathbb{C}^{(-)}_{\gamma \rho } -
      \mathbb{C}^{(0)}_{\mu \omega }\mathbb{C}^{(0)}_{\nu \gamma }\mathbb{C}^{(0)}_{\delta \rho } +
      \mathbb{C}^{(0)}_{\mu \omega }\mathbb{C}^{(0)}_{\nu \rho }\mathbb{C}^{(0)}_{\delta \gamma }. \nonumber\\
\end{eqnarray}

The second class $O^{(1)}$ corresponds to contractions between one pair of operators in $\{ \hat{c}^\dag_{\mu} \hat{c}^\dag_{\nu} \hat{c}^\dag_{\delta} \hat{c}_{\gamma} \hat{c}_{\rho} \hat{c}_{\omega} \}$ and one pair of operators in $\{\hat z_i, \ 1 \leqslant i \leqslant 2N \}$, multiplied by contractions among the left two pairs in $\{ \hat{c}^\dag_{\mu} \hat{c}^\dag_{\nu} \hat{c}^\dag_{\delta} \hat{c}_{\gamma} \hat{c}_{\rho} \hat{c}_{\omega} \}$ then multiplied by contractions among the left $N-1$ pairs in $\{\hat z_i, \ 1 \leqslant i \leqslant 2N \}$, i.e.,
\begin{eqnarray} \label{O1} 
  O^{(1)} = \sum_{ij}^{2N} \mathbb{W}_{ij}^{(1)} (-1)^{i+j} \alpha_{ij} \text{Pf}(\mathbb{S}\{i,j\}) ,
\end{eqnarray}
where
\begin{eqnarray} 
  \mathbb{W}_{ij}^{(1)} = \sum_{\mu \nu \delta \omega\rho\gamma} W_{\mu\nu\delta \omega\rho\gamma} 
  \mathbb{D}_{\mu\nu\delta \gamma\rho\omega}^{ij},
\end{eqnarray}
\begin{eqnarray} 
  \mathbb{D}_{\mu\nu\delta \gamma\rho\omega}^{ij}
  &=& \mathbb{S}_{\mu i}^{(+)} \mathbb{S}_{\nu j}^{(+)} 
      \left( \mathbb{C}^{(0)}_{\delta \gamma} \mathbb{C}^{(-)}_{\rho \omega} - \mathbb{C}^{(0)}_{\delta \rho} \mathbb{C}^{(-)}_{\gamma \omega} +
      \mathbb{C}^{(0)}_{\delta \omega} \mathbb{C}^{(-)}_{\gamma \rho} \right)   \nonumber\\
  &+& \mathbb{S}_{\mu i}^{(+)} \mathbb{S}_{\delta j}^{(+)}
      \left( - \mathbb{C}^{(0)}_{\nu \gamma} \mathbb{C}^{(-)}_{\rho \omega} + \mathbb{C}^{(0)}_{\nu \rho } \mathbb{C}^{(-)}_{\gamma \omega } -
      \mathbb{C}^{(0)}_{\nu \omega }\mathbb{C}^{(-)}_{\gamma \rho } \right) \nonumber\\
  &+& \mathbb{S}_{\mu i}^{(+)} \mathbb{S}_{\gamma j}^{(-)} 
      \left( \mathbb{C}^{(+)}_{\nu \delta }\mathbb{C}^{(-)}_{\rho \omega }-\mathbb{C}^{(0)}_{\nu \rho }\mathbb{C}^{(0)}_{\delta \omega } + 
      \mathbb{C}^{(0)}_{\nu \omega }\mathbb{C}^{(0)}_{\delta \rho } \right)  \nonumber\\
  &+& \mathbb{S}_{\mu i}^{(+)} \mathbb{S}_{\rho j}^{(-)} 
      \left( -\mathbb{C}^{(+)}_{\nu \delta }\mathbb{C}^{(-)}_{\gamma \omega }+\mathbb{C}^{(0)}_{\nu \gamma }\mathbb{C}^{(0)}_{\delta \omega } -
      \mathbb{C}^{(0)}_{\nu \omega }\mathbb{C}^{(0)}_{\delta \gamma } \right) \nonumber\\
  &+& \mathbb{S}_{\mu i}^{(+)}\mathbb{S}_{\omega j}^{(-)} 
      \left( \mathbb{C}^{(+)}_{\nu \delta }\mathbb{C}^{(-)}_{\gamma \rho }-\mathbb{C}^{(0)}_{\nu \gamma }\mathbb{C}^{(0)}_{\delta \rho } +
      \mathbb{C}^{(0)}_{\nu \rho }\mathbb{C}^{(0)}_{\delta \gamma } \right) \nonumber\\
  &+& \mathbb{S}_{\nu i}^{(+)} \mathbb{S}_{\delta j}^{(+)} 
      \left(\mathbb{C}^{(0)}_{\mu \gamma }\mathbb{C}^{(-)}_{\rho \omega }-\mathbb{C}^{(0)}_{\mu \rho }\mathbb{C}^{(-)}_{\gamma \omega } + 
      \mathbb{C}^{(0)}_{\mu \omega }\mathbb{C}^{(-)}_{\gamma \rho } \right) \nonumber\\
  &+& \mathbb{S}_{\nu i}^{(+)} \mathbb{S}_{\gamma j}^{(-)} 
      \left(-\mathbb{C}^{(+)}_{\mu \delta }\mathbb{C}^{(-)}_{\rho \omega }+\mathbb{C}^{(0)}_{\mu \rho }\mathbb{C}^{(0)}_{\delta \omega } - 
      \mathbb{C}^{(0)}_{\mu \omega }\mathbb{C}^{(0)}_{\delta \rho } \right) \nonumber\\
  &+& \mathbb{S}_{\nu i}^{(+)} \mathbb{S}_{\rho j}^{(-)} 
      \left(\mathbb{C}^{(+)}_{\mu \delta }\mathbb{C}^{(-)}_{\gamma \omega }-\mathbb{C}^{(0)}_{\mu \gamma }\mathbb{C}^{(0)}_{\delta \omega } + 
      \mathbb{C}^{(0)}_{\mu \omega }\mathbb{C}^{(0)}_{\delta \gamma } \right) \nonumber\\
  &+& \mathbb{S}_{\nu i}^{(+)} \mathbb{S}_{\omega j}^{(-)} 
      \left(-\mathbb{C}^{(+)}_{\mu \delta }\mathbb{C}^{(-)}_{\gamma \rho }+\mathbb{C}^{(0)}_{\mu \gamma }\mathbb{C}^{(0)}_{\delta \rho } -
      \mathbb{C}^{(0)}_{\mu \rho }\mathbb{C}^{(0)}_{\delta \gamma } \right) \nonumber\\
  &+& \mathbb{S}_{\delta i}^{(+)} \mathbb{S}_{\gamma j}^{(-)} 
      \left( \mathbb{C}^{(+)}_{\mu \nu }\mathbb{C}^{(-)}_{\rho \omega }-\mathbb{C}^{(0)}_{\mu \rho }\mathbb{C}^{(0)}_{\nu \omega } + 
      \mathbb{C}^{(0)}_{\mu \omega }\mathbb{C}^{(0)}_{\nu \rho } \right) \nonumber\\
  &+& \mathbb{S}_{\delta i}^{(+)} \mathbb{S}_{\rho j}^{(-)} 
      \left(-\mathbb{C}^{(+)}_{\mu \nu }\mathbb{C}^{(-)}_{\gamma \omega }+\mathbb{C}^{(0)}_{\mu \gamma }\mathbb{C}^{(0)}_{\nu \omega } - 
      \mathbb{C}^{(0)}_{\mu \omega }\mathbb{C}^{(0)}_{\nu \gamma } \right) \nonumber\\
  &+& \mathbb{S}_{\delta i}^{(+)} \mathbb{S}_{\omega j}^{(-)} 
      \left(\mathbb{C}^{(+)}_{\mu \nu }\mathbb{C}^{(-)}_{\gamma \rho }-\mathbb{C}^{(0)}_{\mu \gamma }\mathbb{C}^{(0)}_{\nu \rho } + 
      \mathbb{C}^{(0)}_{\mu \rho }\mathbb{C}^{(0)}_{\nu \gamma } \right) \nonumber\\
  &+& \mathbb{S}_{\gamma i}^{(-)} \mathbb{S}_{\rho j}^{(-)} 
      \left(\mathbb{C}^{(+)}_{\mu \nu }\mathbb{C}^{(0)}_{\delta \omega }-\mathbb{C}^{(+)}_{\mu \delta }\mathbb{C}^{(0)}_{\nu \omega } + 
      \mathbb{C}^{(0)}_{\mu \omega }\mathbb{C}^{(+)}_{\nu \delta } \right) \nonumber\\
  &+& \mathbb{S}_{\gamma i}^{(-)} \mathbb{S}_{\omega j}^{(-)} 
      \left(-\mathbb{C}^{(+)}_{\mu \nu }\mathbb{C}^{(0)}_{\delta \rho } + \mathbb{C}^{(+)}_{\mu \delta }\mathbb{C}^{(0)}_{\nu \rho } - 
      \mathbb{C}^{(0)}_{\mu \rho }\mathbb{C}^{(+)}_{\nu \delta } \right) \nonumber\\
  &+& \mathbb{S}_{\rho i}^{(-)} \mathbb{S}_{\omega j}^{(-)} 
      \left(\mathbb{C}^{(+)}_{\mu \nu }\mathbb{C}^{(0)}_{\delta \gamma }-\mathbb{C}^{(+)}_{\mu \delta }\mathbb{C}^{(0)}_{\nu \gamma } + 
      \mathbb{C}^{(0)}_{\mu \gamma }\mathbb{C}^{(+)}_{\nu \delta } \right). \nonumber\\
\end{eqnarray}
In Eq. (\ref{O1}) $\mathbb{S}\{i,j\}$ is defined as a sub-matrix of $\mathbb{S}$ with its $i^{\text{th}}$ and $j^{\text{th}}$ rows and columns removed, the phase $(-1)^{i+j}$ and the permutation phase $\alpha_{ij}$ come from the contractions regarding $\hat z_i$ and $\hat z_j$, in which $\alpha_{ij}=1$ when $i<j$ and $\alpha_{ij}=-1$ when $i>j$. 

Similarly, the third class $O^{(2)}$ corresponds to contractions between two pairs of operators in $\{ \hat{c}^\dag_{\mu} \hat{c}^\dag_{\nu} \hat{c}^\dag_{\delta} \hat{c}_{\gamma} \hat{c}_{\rho} \hat{c}_{\omega} \}$ and two pairs of operators in $\{\hat z_i, \ 1 \leqslant i \leqslant 2N \}$, multiplied by contractions among the left one pair in $\{ \hat{c}^\dag_{\mu} \hat{c}^\dag_{\nu} \hat{c}^\dag_{\delta} \hat{c}_{\gamma} \hat{c}_{\rho} \hat{c}_{\omega} \}$ then multiplied by contractions among the left $N-2$ pairs in $\{\hat z_i, \ 1 \leqslant i \leqslant 2N \}$, i.e.,
\begin{eqnarray} \label{O2} 
  O^{(2)} = \sum_{ijkl}^{2N} \mathbb{W}_{ijkl}^{(2)} (-1)^{i+j+k+l} \alpha_{ijkl} \text{Pf} (\mathbb{S}\{i,j,k,l\}) ,
\end{eqnarray}
where 
\begin{eqnarray} 
  \mathbb{W}_{ijkl}^{(2)} = 
  \sum_{\mu\nu\delta \omega\rho\gamma} W_{\mu\nu\delta\omega\rho\gamma} \mathbb{E}_{\mu \nu\delta\gamma\rho\omega}^{ijkl},
\end{eqnarray}
\begin{eqnarray} 
  \mathbb{E}_{\mu \nu\delta\gamma\rho\omega }^{ijkl}
  &=& \mathbb{S}_{\mu i}^{(+)} \mathbb{S}_{\nu j}^{(+)} \mathbb{S}_{\delta k}^{(+)} \mathbb{S}_{\gamma l}^{(-)} \mathbb{C}_{\rho \omega }^{(-)} -
      \mathbb{S}_{\mu i}^{(+)} \mathbb{S}_{\nu j}^{(+)} \mathbb{S}_{\delta k}^{(+)} \mathbb{S}_{\rho l}^{(-)} \mathbb{C}_{\gamma \omega }^{(-)} 
      \nonumber\\
  &+& \mathbb{S}_{\mu i}^{(+)} \mathbb{S}_{\nu j}^{(+)} \mathbb{S}_{\delta k}^{(+)} \mathbb{S}_{\omega l}^{(-)} \mathbb{C}_{\gamma \rho }^{(-)} +
      \mathbb{S}_{\mu i}^{(+)} \mathbb{S}_{\nu j}^{(+)} \mathbb{S}_{\gamma k}^{(-)} \mathbb{S}_{\rho l}^{(-)} \mathbb{C}_{\delta \omega }^{(0)} 
      \nonumber\\
  &-& \mathbb{S}_{\mu i}^{(+)} \mathbb{S}_{\nu j}^{(+)} \mathbb{S}_{\gamma k}^{(-)} \mathbb{S}_{\omega l}^{(-)} \mathbb{C}_{\delta \rho }^{(0)} +
      \mathbb{S}_{\mu i}^{(+)} \mathbb{S}_{\nu j}^{(+)} \mathbb{S}_{\rho k}^{(-)} \mathbb{S}_{\omega l}^{(-)} \mathbb{C}_{\delta \gamma }^{(0)} 
      \nonumber\\
  &-& \mathbb{S}_{\mu i}^{(+)} \mathbb{S}_{\delta j}^{(+)} \mathbb{S}_{\gamma k}^{(-)} \mathbb{S}_{\rho l}^{(-)} \mathbb{C}_{\nu \omega }^{(0)} +
      \mathbb{S}_{\mu i}^{(+)} \mathbb{S}_{\omega j}^{(-)} \mathbb{S}_{\delta k}^{(+)} \mathbb{S}_{\gamma l}^{(-)} \mathbb{C}_{\nu \rho }^{(0)}
      \nonumber\\
  &-& \mathbb{S}_{\mu i}^{(+)} \mathbb{S}_{\delta j}^{(+)} \mathbb{S}_{\rho k}^{(-)} \mathbb{S}_{\omega l}^{(-)} \mathbb{C}_{\nu \gamma }^{(0)} +
      \mathbb{S}_{\mu i}^{(+)} \mathbb{S}_{\gamma j}^{(-)} \mathbb{S}_{\rho k}^{(-)} \mathbb{S}_{\omega l}^{(-)} \mathbb{C}_{\nu \delta }^{(+)} 
      \nonumber\\
  &+& \mathbb{S}_{\nu i}^{(+)} \mathbb{S}_{\delta j}^{(+)} \mathbb{S}_{\gamma k}^{(-)} \mathbb{S}_{\rho l}^{(-)} \mathbb{C}_{\mu \omega }^{(0)} -
      \mathbb{S}_{\nu i}^{(+)} \mathbb{S}_{\omega j}^{(-)} \mathbb{S}_{\delta k}^{(+)} \mathbb{S}_{\gamma l}^{(-)} \mathbb{C}_{\mu \rho }^{(0)} 
      \nonumber\\
  &+& \mathbb{S}_{\nu i}^{(+)} \mathbb{S}_{\delta j}^{(+)} \mathbb{S}_{\rho k}^{(-)} \mathbb{S}_{\omega l}^{(-)} \mathbb{C}_{\mu \gamma }^{(0)} -
      \mathbb{S}_{\nu i}^{(+)} \mathbb{S}_{\gamma j}^{(-)} \mathbb{S}_{\rho k}^{(-)} \mathbb{S}_{\omega l}^{(-)} \mathbb{C}_{\mu \delta }^{(+)}
      \nonumber\\
  &+& \mathbb{S}_{\delta i}^{(+)} \mathbb{S}_{\gamma j}^{(-)} \mathbb{S}_{\rho k}^{(-)} \mathbb{S}_{\omega l}^{(-)} \mathbb{C}_{\mu \nu }^{(+)}.
\end{eqnarray}
In Eq. (\ref{O2}) $\mathbb{S}\{i, j, k, l\}$ is defined as a sub-matrix of $\mathbb{S}$ with its $i^{\text{th}}$, $j^{\text{th}}$, $k^{\text{th}}$ and $l^{\text{th}}$ rows and columns removed, the phase $(-1)^{i+j+k+l}$ and the permutation phase $\alpha_{ijkl}$ come from the contractions regarding $\hat z_i$, $\hat z_j$, $\hat z_k$ and $\hat z_l$, where $\alpha_{ijkl}=1$ when $i<j<k<l$ and changes its sign one time for each permutation among the indices $i, j, k, l$.  

The last class $O^{(3)}$ reflects the last possible contraction way, i.e., each of the (all the) three pairs of operators in $\{ \hat{c}^\dag_{\mu} \hat{c}^\dag_{\nu} \hat{c}^\dag_{\delta} \hat{c}_{\gamma} \hat{c}_{\rho} \hat{c}_{\omega} \}$ are contracted with each of the three pairs of operators in $\{\hat z_i, \ 1 \leqslant i \leqslant 2N \}$, multiplied by contractions among the left $N-3$ pairs in $\{\hat z_i, \ 1 \leqslant i \leqslant 2N \}$, for this way one has,
\begin{eqnarray} \label{O3}  
  O^{(3)} &=& \sum_{ijklmn}^{2N} \mathbb{W}^{(3)}_{ijklmn} (-1)^{i+j+k+l+m+n} \alpha_{ijklmn} \nonumber \\
          & & \qquad \quad \times \text{Pf}(\mathbb{S}\{i,j,k,l,m,n\}),
\end{eqnarray}
where
\begin{eqnarray} 
  \mathbb{W}_{ijklmn}^{(3)} = \sum_{\mu\nu\delta \omega\rho\gamma} 
  W_{\mu\nu\delta\omega\rho\gamma} \mathbb{F}_{\mu \nu\delta\gamma\rho\omega}^{ijklmn},
\end{eqnarray}
\begin{eqnarray}
  \mathbb{F}_{\mu\nu\delta\gamma\rho\omega}^{ijklmn} = 
  \mathbb{S}_{\mu i}^{(+)} \mathbb{S}_{\nu j}^{(+)} \mathbb{S}_{\delta k}^{(+)}
  \mathbb{S}_{\gamma l}^{(-)} \mathbb{S}_{\rho m}^{(-)} \mathbb{S}_{\omega n}^{(-)}.
\end{eqnarray}
Here the matrix $\mathbb{S}\{i,j,k,l,m,n\}$ and the phases $(-1)^{i+j+k+l+m+n}, \alpha_{ijklmn}$ are defined as the similar way as the $O^{(2)}$ case.

From Eqs. (\ref{I3_sum}, \ref{O0}, \ref{O1}, \ref{O2}, \ref{O3}) we can get a Pfaffian formula for the three-body operator matrix elements as,
\begin{widetext}
\begin{eqnarray} \label{I_3_final_a} 
  \frac{I_3}{\langle \Phi | \Phi' \rangle}
  &=& O^{(0)}+O^{(1)}+O^{(2)}+O^{(3)} \nonumber\\
  &=& W_{0} \text{Pf} (\mathbb{S}) + \sum_{ij}^{2N} \mathbb{W}_{ij}^{(1)} (-1)^{i+j} \alpha_{ij} \text{Pf}(\mathbb{S}\{i,j\}) 
      + \sum_{ijkl}^{2N} \mathbb{W}_{ijkl}^{(2)} (-1)^{i+j+k+l} \alpha_{ijkl} \text{Pf} (\mathbb{S}\{i,j,k,l\}) \nonumber\\
  & & + \sum_{ijklmn}^{2N} \mathbb{W}^{(3)}_{ijklmn} (-1)^{i+j+k+l+m+n} \alpha_{ijklmn} \text{Pf}(\mathbb{S}\{i,j,k,l,m,n\}) .
\end{eqnarray}
\end{widetext}

Such an algorithm in Eq. (\ref{I_3_final_a}) can also be obtained by adopting the expansion property of Pfaffian with respect to six neighboring rows (see Appendix). 

In most cases of practical applications, the norm overlaps are nonzero, so that we have $\text{Pf}(\mathbb{S}) \neq 0$ from Eqs. (\ref{kernels}, \ref{Pf_norm}). For these cases the inverse of the matrix $\mathbb{S}$ would exist and we can then get a more compact and efficient expression for $I_3$ by applying the following Pfaffian identity (the Pfaffian version of the Lewis Carroll formula) which is derived by Mizusaki and Oi (see the Eq. 49 of Ref. \cite{Mizusaki_2012_PLB})
\begin{eqnarray} \label{Mizusaki_49}  
  \text{Pf} \left( \mathbb{X} \right) \text{Pf} \left[ \left(\mathbb{X}^{-1} \right)_{I} \right] 
  = (-1)^{|I|} \text{Pf} \left( \mathbb{X}_{\bar{I}} \right),
\end{eqnarray}
which holds for any skew-symmetric matrix $\mathbb{X}$. Let the matrix $\mathbb{X}$ has $2N \times 2N$ elements as the matrix $\mathbb{S}$ in the above discussions, and employ $[2N] \equiv \{1, 2, 3, \cdots , 2N\}$ to denote a set of integers which correspond to the numbers of rows and columns of the matrix $\mathbb{X}$. We divide $[2N]$ in two groups, with $I \equiv \{ i_{1}, i_{2}, i_{3}, \cdots , i_{2n} \}$ denote a set of indices which corresponds to a subset of $[2N]$ with $1 \leqslant \{ i_{1}, i_{2}, i_{3}, \cdots , i_{2n} \} \leqslant 2N$, the rest of the indices are denoted as $\bar I = [2N] - I$ meaning the complementary group of $I$ in $[2N]$. In Eq. (\ref{Mizusaki_49}) $|I| = \sum_{k=1}^{2n} i_{k}$, $\mathbb{X}^{-1}$ labels the inverse matrix of $\mathbb X$ and $\mathbb{X}_{I}$ represents a $2n \times 2n$ skew matrix with its matrix elements being expressed as $(\mathbb{X}_{I})_{k,l}=\mathbb{X}_{i_{k},i_{l}}$. The notation $\mathbb{X}_{\bar I}$ labels a $2(N-n)\times2(N-n)$ sub-matrix of the matrix $\mathbb X$ by removing the rows and columns of $\{ i_{1}, i_{2}, i_{3}, \cdots , i_{2n} \}$ from the original matrix $\mathbb X$.

From the Pfaffian identity shown in Eq. (\ref{Mizusaki_49}), the Pfaffians of sub-matrix in Eq. (\ref{I_3_final_a}) can be avoided and the expression of $I_3$ in Eq. (\ref{I_3_final_a}) can by further written as (see Appendix),

\begin{eqnarray} \label{I_3_final_b}  
  \frac{I_3}{\langle \Phi | \Phi' \rangle}
  &=& W_{0} \text{Pf}(\mathbb S) - \text{Tr} \left( \mathbb{W}^{(1)} \mathbb{S}^{-1} \right) \text{Pf}(\mathbb S) \nonumber \\
  &+& \sum_{ijkl}^{2N} \mathbb{W}_{ijkl}^{(2)} \left( \mathbb{S}_{ij}^{-1} \mathbb{S}_{kl}^{-1} - 
      \mathbb{S}_{ik}^{-1} \mathbb{S}_{jl}^{-1} +\mathbb{S}_{il}^{-1} \mathbb{S}_{jk}^{-1} \right) \text{Pf}(\mathbb S)  \nonumber \\
  &+& \sum_{ijklmn}^{2N} \mathbb{W}_{ijklmn}^{(3)} \mathbb{S}_{ijklmn}^{-1} \text{Pf}(\mathbb S) .
\end{eqnarray}
where
\begin{eqnarray}
  \mathbb{S}_{ijklmn}^{-1}
  &=& \mathbb{S}_{ij}^{-1} \mathbb{S}_{kl}^{-1} \mathbb{S}_{mn}^{-1} - \mathbb{S}_{ij}^{-1} \mathbb{S}_{km}^{-1} \mathbb{S}_{ln}^{-1} + 
      \mathbb{S}_{ij}^{-1} \mathbb{S}_{kn}^{-1} \mathbb{S}_{lm}^{-1} \nonumber \\
  &-& \mathbb{S}_{ik}^{-1} \mathbb{S}_{jl}^{-1} \mathbb{S}_{mn}^{-1} + \mathbb{S}_{ik}^{-1} \mathbb{S}_{jm}^{-1} \mathbb{S}_{ln}^{-1} - 
      \mathbb{S}_{ik}^{-1} \mathbb{S}_{jn}^{-1} \mathbb{S}_{lm}^{-1} \nonumber \\
  &+& \mathbb{S}_{il}^{-1} \mathbb{S}_{jk}^{-1} \mathbb{S}_{mn}^{-1} - \mathbb{S}_{il}^{-1} \mathbb{S}_{jm}^{-1} \mathbb{S}_{kn}^{-1} + 
      \mathbb{S}_{il}^{-1} \mathbb{S}_{jn}^{-1} \mathbb{S}_{km}^{-1} \nonumber \\
  &-& \mathbb{S}_{im}^{-1} \mathbb{S}_{jk}^{-1} \mathbb{S}_{ln}^{-1} + \mathbb{S}_{im}^{-1} \mathbb{S}_{jl}^{-1} \mathbb{S}_{kn}^{-1} - 
      \mathbb{S}_{im}^{-1} \mathbb{S}_{jn}^{-1} \mathbb{S}_{kl}^{-1} \nonumber \\
  &+& \mathbb{S}_{in}^{-1} \mathbb{S}_{jk}^{-1} \mathbb{S}_{lm}^{-1} - \mathbb{S}_{in}^{-1} \mathbb{S}_{jl}^{-1} \mathbb{S}_{km}^{-1} + 
      \mathbb{S}_{in}^{-1} \mathbb{S}_{jm}^{-1} \mathbb{S}_{kl}^{-1}. \nonumber \\
\end{eqnarray}

Now let us remark the differences between the calculation of three-body operator matrix elements $I_3$ by Eq. (\ref{3b_ME_b}) directly and the evaluation of $I_3$ by the Pfaffian formula in Eq. (\ref{I_3_final_b}). For the former way, due to the 6-fold loops of $\{\mu\nu\delta \omega\rho\gamma\}$ in large model space (with dimension $M$), as many as $M^6$ Pfaffians of $(2N+6)\times(2N+6)$ matrices need to be calculated numerically, which would turn out to be much time-consuming for large model space as the calculation time of Pfaffian is non-negligible \cite{Pfaffian_code_CPC_2011}. On the other hand, we note the fact that for either the spectroscopy or the transition/decay problems, the norm overlaps in Eqs. (\ref{kernels}a, \ref{kernels}b) should be calculated first before the evaluation of matrix elements of physical operators in Eq. (\ref{kernels}c) (or $I_3$ in the above discussions). This indicates that, as seen from Eq. (\ref{Pf_norm}), the matrix $\mathbb S$, its Pfaffian $\text{Pf}(\mathbb S)$ and its reverse $\mathbb S^{-1}$ have already been prepared and stored to the memory before the evaluation of $I_3$, which means that the Pfaffian formula in Eq. (\ref{I_3_final_b}) should be much efficient than Eq. (\ref{3b_ME_b}) since taking data from memory and then making manipulation is usually much faster than preparing matrices and then calculating their Pfaffians.

More interestingly, one can further reduce the Pfaffian formula in Eq. (\ref{I_3_final_b}) according to the underlying physics in practical applications. As discussed in Eqs. (\ref{I3_sum}, \ref{O0}, \ref{O1}, \ref{O2}, \ref{O3}), the four terms in Eq. (\ref{I_3_final_b}) corresponds to contractions of the three-body operator with no pair, one pair, two pairs and three pairs of operators in the configurations $\{\hat z_i, \ 1 \leqslant i \leqslant 2N \}$, respectively. Therefore, for even-even nuclei, if only the collective degrees of freedom are of interests for studies of low-lying states so that only qp vacua are included in the configuration space (as for most of the current GCM models \cite{GCM_Yao_2010, GCM_Bender_2008, GCM_Tomas}), we would have $N=0$ and only the first term $W_{0} \text{Pf}(\mathbb S)$ survives in Eq. (\ref{I_3_final_b}). Similarly, for low-lying states of odd-mass nuclei, when only the 1-qp configurations \cite{Carlsson_2021_PRL} are considered we then have $N=2$ and only the first two terms in Eq. (\ref{I_3_final_b}) would survive. For an ambitious GCM model that considers both collective and single-particle degrees of freedom taking into account up to 2-qp configurations as in Ref. \cite{GCM_qp_FQChen_2017}, one has $N \leqslant 4$ and does not need to worry about the last term in Eq. (\ref{I_3_final_b}).

Finally, we discuss a rare case in which $\mathbb{S}^{-1}$ does not exist so that Eq. (\ref{I_3_final_b}) is no longer valid. For such case we can get another expression for $I_3$ readily in the similar way as in Ref. \cite{Hu_arXiv} (see the Eqs. (38, 39, 53) of Ref. \cite{Hu_arXiv} for details),
\begin{widetext}
\begin{eqnarray}
  \frac{I_3}{\langle \Phi | \Phi' \rangle}
  &=& W_{0} \text{Pf} (\mathbb{S}) - \sum_{ij}^{2N} (-1)^{i+j+1} \alpha_{ij} \tilde{\mathbb{S}}^{i}_{ij} \text{Pf} (\tilde{\mathbb{S}}^{i}\{i,j\}) 
      +\sum_{ij}^{2N} (-1)^{i+j+1} \alpha_{ij} \sum_{kl}^{2N} (-1)^{k+l+1} \alpha_{kl} \tilde{\mathbb{S}}^{ijk}_{ijkl} 
      \text{Pf} (\tilde{\mathbb{S}}^{ijk} \{i,j,k,l\})  \nonumber \\
  & & \qquad + \sum_{ijklmn}^{2N} \mathbb{W}^{(3)}_{ijklmn} (-1)^{i+j+k+l+m+n} \alpha_{ijklmn} \text{Pf}(\mathbb{S}\{i,j,k,l,m,n\}) \nonumber \\
  &=& W_0 \text{Pf} (\mathbb{S}) - \sum_{i}^{2N} \text{Pf} (\tilde{\mathbb{S}}^{i})  
      +\sum_{ij}^{2N} (-1)^{i+j+1} \alpha_{ij} \sum_k^{2N} \text{Pf}(\tilde{\mathbb{S}}^{ijk}\{i,j\}) \nonumber \\
  & & \qquad + \sum_{ijklmn}^{2N} \mathbb{W}^{(3)}_{ijklmn} (-1)^{i+j+k+l+m+n} \alpha_{ijklmn} \text{Pf}(\mathbb{S}\{i,j,k,l,m,n\}) .
\end{eqnarray}
\end{widetext}
where $\tilde{\mathbb{S}}^{i}$ is constructed in the way that only replacing the $i^{\text{th}}$ row and column of matrix $\mathbb{S}$ by the $i^{\text{th}}$ row of $\mathbb W^{(1)}$, i.e., $\tilde{\mathbb{S}}_{ij}^{i} = \mathbb{W}^{(1)}_{ij}$ and keep the skew-symmetry. The matrix $\tilde{\mathbb S}^{ijk}$ is the same as $\mathbb S$ except for replacing the $k^{\text{th}}$ row and column by the matrix elements of $\mathbb W^{(2)}$, i.e., $\tilde{\mathbb S}^{ijk}_{ijkl} = \mathbb W^{(2)}_{ijkl}$ and keep the skew-symmetry of $\tilde{\mathbb S}^{ijk}$. 


There are some potential applications of the Pfaffian formula for matrix elements of three-body operators in the near future. For beyond-mean-field nuclear models (especially the AMP-based method such as the GCM) with realistic nuclear forces, calculations of matrix elements of general three-body operators are needed if effects beyond the normal-ordering approximations are taken into account \cite{Hebeler_Phys_Rep_2021}. Besides, for nuclear neutrinoless double-$\beta$ decay which is one of the hot topics in modern nuclear physics, the nature of neutrinos can be better explored and understood, provided that the corresponding nuclear matrix elements can be evaluated as precisely as possible. The nuclear matrix elements $M^{0\nu\beta\beta}$ can be written as,
\begin{eqnarray}
  M^{0\nu\beta\beta} = \left\langle \Psi_F \left| \hat O ^{0\nu\beta\beta} \right| \Psi_I \right\rangle.
\end{eqnarray}
from which one can see that there are two sources of uncertainties for $M^{0\nu\beta\beta}$, the one from nuclear many-body wave functions of the parent $| \Psi_I \rangle$ and daughter $| \Psi_F \rangle$ nuclei, and the one from the decay operator $\hat O ^{0\nu\beta\beta}$. The uncertainty in the former can be reduced to a large extent by taking into account as many correlations as possible in the nuclear wave function, such as the shape fluctuations \cite{Tomas_2010_PRL_0vbb, Yao_2015_PRC_0vbb}, pairing fluctuations \cite{Tomas_2013_PRL_pairing_0vbb, GCM_Nobuo_2014}, qp excitation \cite{Y_K_Wang_PRC_2021_0vbb} etc. for which the GCM method serves as the optimal candidate model. The uncertainty from the decay operator can be reduced effectively by studying the roles of chiral two-body currents \cite{LJWang_current_2018_Rapid}, which would lead to three-body and even four-body decay operators. Therefore, to provide reliable $M^{0\nu\beta\beta}$ with minimum uncertainties, matrix elements of three-body decay operators in the GCM methods with/without qp configurations are indispensable, for which our Pfaffian formula in Eq. (\ref{I_3_final_b}) is expected to be useful.

\section{\label{sec:sum}summary}

To summarize, we present a compact and efficient Pfaffian algorithm for evaluation of matrix elements of any three-body operators in beyond-mean-field nuclear models such as the GCM method, for cases with or without qp configurations. Further optimization of the Pfaffian algorithm in practical nuclear-structure problems is discussed for cases such as the low-lying states of even-even or odd-mass nuclei. Potential applications of the algorithm in nuclear physics are explored, including developing AMP or GCM based models with realistic nuclear forces and reducing the uncertainties in nuclear matrix elements of neutrinoless double beta decays.

\begin{acknowledgments}
  L.J.W. would like to thank F.-Q. Chen, Q. L. Hu and J. M. Yao for many discussions. This work is supported by the National Natural Science Foundation of China (Grant Nos. 11905175, 11875225, and U1932206), and by the Venture $\&$ Innovation Support Program for Chongqing Overseas Returnees (with Grant No. cx2019056).
\end{acknowledgments}

\appendix*

\section{} \label{app}

First, we show that the Eq. (\ref{I_3_final_a}) can be obtained equivalently by the expansion property of the Pfaffian with respect to six neighboring rows or columns which can be derived in the similar way as in Refs. \cite{Hu_2014_PLB, Hu_arXiv} based on the Lemma 2.3 of Ref. \cite{Ishikawa_Pfaffian_1999}. The derivation is complicated and tedious so that we only provide the conclusion in the following. 

The expansion property of the Pfaffian for a matrix $X$ with respect to the neighboring $i_0^{\text{th}}$, $j_0^{\text{th}}$, $k_0^{\text{th}}$, $l_0^{\text{th}}$, $m_0^{\text{th}}$, and $n_0^{\text{th}}$ rows reads as,
\begin{widetext}
\begin{eqnarray} \label{expansion_Pf}
  \text{Pf}(X) 
  &=& Y_{i_{0} j_{0} k_{0} l_{0} m_{0} n_{0} } \text{Pf} (X\{i_{0}, j_{0}, k_{0}, l_{0}, m_{0}, n_{0}\}) \nonumber \\
  &+& \sum_{ij} (-1)^{i+j} \alpha_{ij} Z_{i_{0} j_{0} k_{0} l_{0} m_{0} n_{0}}^{ij} \text{Pf} (X\{i_{0}, j_{0}, k_{0}, l_{0}, m_{0}, n_{0}, i, j \})
      \nonumber \\
  &+& \sum_{ijkl} (-1)^{i+j+k+l} \alpha_{ijkl} W_{i_{0} j_{0} k_{0} l_{0} m_{0} n_{0}}^{ijkl} \text{Pf} (X\{i_{0}, j_{0}, k_{0}, l_{0}, m_{0}, n_{0}, i, j, k, l \}) \nonumber \\
  &+& \sum_{ijklmn} (-1)^{i+j+k+l+m+n} \alpha_{ijklmn} U_{i_{0} j_{0} k_{0} l_{0} m_{0} n_{0}}^{ijklmn} \text{Pf} 
      (X\{i_{0}, j_{0}, k_{0}, l_{0}, m_{0}, n_{0}, i, j, k, l, m, n \})  ,
\end{eqnarray}
where
\begin{eqnarray}
  Y_{i_{0}j_{0}k_{0}l_{0}m_{0}n_{0}} 
  &=&  X_{i_{0} j_{0} } X_{k_{0} l_{0} }X_{m_{0} n_{0} }-X_{i_{0} j_{0} }X_{k_{0} m_{0} }X_{l_{0} n_{0} }+X_{i_{0} j_{0} }X_{k_{0} n_{0} }X_{l_{0} m_{0} } \nonumber \\
  &-& X_{i_{0} k_{0} }X_{j_{0} l_{0} }X_{m_{0} n_{0} }+X_{i_{0} k_{0} }X_{j_{0} m_{0} }X_{l_{0} n_{0} }-X_{i_{0} k_{0} }X_{j_{0} n_{0} }X_{l_{0} m_{0} } \nonumber \\
  &+& X_{i_{0} l_{0} }X_{j_{0} k_{0} }X_{m_{0} n_{0} }-X_{i_{0} l_{0} }X_{j_{0} m_{0} }X_{k_{0} n_{0} }+X_{i_{0} l_{0} }X_{j_{0} n_{0} }X_{k_{0} m_{0} } \nonumber \\
  &-& X_{i_{0} m_{0} }X_{j_{0} k_{0} }X_{l_{0} n_{0} }+X_{i_{0} m_{0} }X_{j_{0} l_{0} }X_{k_{0} n_{0} }-X_{i_{0} m_{0} }X_{j_{0} n_{0} }X_{k_{0} l_{0} } \nonumber \\
  &+& X_{i_{0} n_{0} }X_{j_{0} k_{0} }X_{l_{0} m_{0} }-X_{i_{0} n_{0} }X_{j_{0} l_{0} }X_{k_{0} m_{0} }+X_{i_{0} n_{0} }X_{j_{0} m_{0} }X_{k_{0} l_{0} } ,
\end{eqnarray}
\begin{eqnarray}
  Z_{i_{0}j_{0}k_{0}l_{0}m_{0}n_{0}}^{ij} 
  &=& X_{i_{0} i}X_{j_{0} j}X_{k_{0} l_{0} }X_{m_{0} n_{0} }-X_{i_{0} i}X_{j_{0} j}X_{k_{0} m_{0} }X_{l_{0} n_{0} }+X_{i_{0} i}X_{j_{0} j}X_{k_{0} n_{0} }X_{l_{0} m_{0} } \nonumber \\
  &-& X_{i_{0} i}X_{k_{0} j}X_{j_{0} l_{0} }X_{m_{0} n_{0} }+X_{i_{0} i}X_{k_{0} j}X_{j_{0} m_{0} }X_{l_{0} n_{0} }-X_{i_{0} i}X_{k_{0} j}X_{j_{0} n_{0} }X_{l_{0} m_{0} } \nonumber \\
  &+& X_{i_{0} i}X_{l_{0} j}X_{j_{0} k_{0} }X_{m_{0} n_{0} }-X_{i_{0} i}X_{l_{0} j}X_{j_{0} m_{0} }X_{k_{0} n_{0} }+X_{i_{0} i}X_{l_{0} j}X_{j_{0} n_{0} }X_{k_{0} m_{0} } \nonumber \\
  &-& X_{i_{0} i}X_{m_{0} j}X_{j_{0} k_{0} }X_{l_{0} n_{0} }+X_{i_{0} i}X_{m_{0} j}X_{j_{0} l_{0} }X_{k_{0} n_{0} }-X_{i_{0} i}X_{m_{0} j}X_{j_{0} n_{0} }X_{k_{0} l_{0} } \nonumber \\
  &+& X_{i_{0} i}X_{n_{0} j}X_{j_{0} k_{0} }X_{l_{0} m_{0} }-X_{i_{0} i}X_{n_{0} j}X_{j_{0} l_{0} }X_{k_{0} m_{0} }+X_{i_{0} i}X_{n_{0} j}X_{j_{0} m_{0} }X_{k_{0} l_{0} } \nonumber \\
  &+& X_{j_{0} i}X_{k_{0} j}X_{i_{0} l_{0} }X_{m_{0} n_{0} }-X_{j_{0} i}X_{k_{0} j}X_{i_{0} m_{0} }X_{l_{0} n_{0} }+X_{j_{0} i}X_{k_{0} j}X_{i_{0} n_{0} }X_{l_{0} m_{0} } \nonumber \\
  &-& X_{j_{0} i}X_{l_{0} j}X_{i_{0} k_{0} }X_{m_{0} n_{0} }+X_{j_{0} i}X_{l_{0} j}X_{i_{0} m_{0} }X_{k_{0} n_{0} }-X_{j_{0} i}X_{l_{0} j}X_{i_{0} n_{0} }X_{k_{0} m_{0} } \nonumber \\
  &+& X_{j_{0} i}X_{m_{0} j}X_{i_{0} k_{0} }X_{l_{0} n_{0} }-X_{j_{0} i}X_{m_{0} j}X_{i_{0} l_{0} }X_{k_{0} n_{0} }+X_{j_{0} i}X_{m_{0} j}X_{i_{0} n_{0} }X_{k_{0} l_{0} } \nonumber \\
  &-& X_{j_{0} i}X_{n_{0} j}X_{i_{0} k_{0} }X_{l_{0} m_{0} }+X_{j_{0} i}X_{n_{0} j}X_{i_{0} l_{0} }X_{k_{0} m_{0} }-X_{j_{0} i}X_{n_{0} j}X_{i_{0} m_{0} }X_{k_{0} l_{0} } \nonumber \\
  &+& X_{k_{0} i}X_{l_{0} j}X_{i_{0} j_{0} }X_{m_{0} n_{0} }-X_{k_{0} i}X_{l_{0} j}X_{i_{0} m_{0} }X_{j_{0} n_{0} }+X_{k_{0} i}X_{l_{0} j}X_{i_{0} n_{0} }X_{j_{0} m_{0} } \nonumber \\
  &-& X_{k_{0} i}X_{m_{0} j}X_{i_{0} j_{0} }X_{l_{0} n_{0} }+X_{k_{0} i}X_{m_{0} j}X_{i_{0} l_{0} }X_{j_{0} n_{0} }-X_{k_{0} i}X_{m_{0} j}X_{i_{0} n_{0} }X_{j_{0} l_{0} } \nonumber \\
  &+& X_{k_{0} i}X_{n_{0} j}X_{i_{0} j_{0} }X_{l_{0} m_{0} }-X_{k_{0} i}X_{n_{0} j}X_{i_{0} l_{0} }X_{j_{0} m_{0} }+X_{k_{0} i}X_{n_{0} j}X_{i_{0} m_{0} }X_{j_{0} l_{0} } \nonumber \\
  &+& X_{l_{0} i}X_{m_{0} j}X_{i_{0} j_{0} }X_{k_{0} n_{0} }-X_{l_{0} i}X_{m_{0} j}X_{i_{0} k_{0} }X_{j_{0} n_{0} }+X_{l_{0} i}X_{m_{0} j}X_{i_{0} n_{0} }X_{j_{0} k_{0} } \nonumber \\
  &-& X_{l_{0} i}X_{n_{0} j}X_{i_{0} j_{0} }X_{k_{0} m_{0} }+X_{l_{0} i}X_{n_{0} j}X_{i_{0} k_{0} }X_{j_{0} m_{0} }-X_{l_{0} i}X_{n_{0} j}X_{i_{0} m_{0} }X_{j_{0} k_{0} } \nonumber \\
  &+& X_{m_{0} i}X_{n_{0} j}X_{i_{0} j_{0} }X_{k_{0} l_{0} }-X_{m_{0} i}X_{n_{0} j}X_{i_{0} k_{0} }X_{j_{0} l_{0} }+X_{m_{0} i}X_{n_{0} j}X_{i_{0} l_{0} }X_{j_{0} k_{0} } ,
\end{eqnarray}
\begin{eqnarray}
  W_{i_{0}j_{0}k_{0}l_{0}m_{0}n_{0}}^{ijkl} 
  &=& X_{i_{0} i}X_{j_{0} j}X_{k_{0} k}X_{l_{0} l}X_{m_{0} n_{0} }- X_{i_{0} i}X_{j_{0} j}X_{k_{0} k}X_{m_{0} l}X_{l_{0} n_{0} } \nonumber \\
  &+& X_{i_{0} i}X_{j_{0} j}X_{k_{0} k}X_{n_{0} l}X_{l_{0} m_{0} } +X_{i_{0} i}X_{j_{0} j}X_{l_{0} k}X_{m_{0} l}X_{k_{0} n_{0} } \nonumber \\
  &-& X_{i_{0} i}X_{j_{0} j}X_{l_{0} k}X_{n_{0} l}X_{k_{0} m_{0} } +X_{i_{0} i}X_{j_{0} j}X_{m_{0} k}X_{n_{0} l}X_{k_{0} l_{0} } \nonumber \\
  &-& X_{i_{0} i}X_{k_{0} j}X_{l_{0} k}X_{m_{0} l}X_{j_{0} n_{0} } +X_{i_{0} i}X_{n_{0} j}X_{k_{0} k}X_{l_{0} l}X_{j_{0} m_{0} } \nonumber \\
  &-& X_{i_{0} i}X_{k_{0} j}X_{m_{0} k}X_{n_{0} l}X_{j_{0} l_{0} } +X_{i_{0} i}X_{l_{0} j}X_{m_{0} k}X_{n_{0} l}X_{j_{0} k_{0} } \nonumber \\
  &+& X_{j_{0} i}X_{k_{0} j}X_{l_{0} k}X_{m_{0} l}X_{i_{0} n_{0} } -X_{j_{0} i}X_{n_{0} j}X_{k_{0} k}X_{l_{0} l}X_{i_{0} m_{0} } \nonumber \\
  &+& X_{j_{0} i}X_{k_{0} j}X_{m_{0} k}X_{n_{0} l}X_{i_{0} l_{0} } -X_{j_{0} i}X_{l_{0} j}X_{m_{0} k}X_{n_{0} l}X_{i_{0} k_{0} } \nonumber \\
  &+& X_{k_{0} i}X_{l_{0} j}X_{m_{0} k}X_{n_{0} l}X_{i_{0} j_{0} } ,
\end{eqnarray}
\begin{eqnarray}
  U_{i_{0}j_{0}k_{0}l_{0}m_{0}n_{0}}^{ijklmn} = X_{i_{0} i}X_{j_{0} j}X_{k_{0} k}X_{l_{0} l}X_{m_{0} m}X_{n_{0} n} .
\end{eqnarray}
from Eqs. (\ref{3b_ME_b}, \ref{expansion_Pf}) one can then obtain Eq. (\ref{I_3_final_a}) readily. 

Secondly we show that from Eq. (\ref{Mizusaki_49}) we get,
\begin{subequations}
\begin{eqnarray}
  \text{Pf} ( \mathbb{S} \{i,j\} ) 
  &=& (-1)^{i+j} \alpha_{ij} \mathbb{S}_{ij}^{-1} \text{Pf} (\mathbb{S}), \\
  \text{Pf} (\mathbb{S}\{i,j,k,l\})
  &=& (-1)^{i+j+k+l} \alpha_{ijkl} \left( \mathbb{S}_{ij}^{-1} \mathbb{S}_{kl}^{-1} - \mathbb{S}_{ik}^{-1} \mathbb{S}_{jl}^{-1} + \mathbb{S}_{il}^{-1} \mathbb{S}_{jk}^{-1} \right) \text{Pf} (\mathbb{S}), \\
  \text{Pf} (\mathbb{S}\{i,j,k,l,m,n\})
  &=& (-1)^{i+j+k+l+m+n} \alpha_{ijklmn} \Big( 
        \mathbb{S}_{ij}^{-1} \mathbb{S}_{kl}^{-1} \mathbb{S}_{mn}^{-1}
      - \mathbb{S}_{ij}^{-1} \mathbb{S}_{km}^{-1} \mathbb{S}_{ln}^{-1}
      + \mathbb{S}_{ij}^{-1} \mathbb{S}_{kn}^{-1} \mathbb{S}_{lm}^{-1} 
      - \mathbb{S}_{ik}^{-1} \mathbb{S}_{jl}^{-1} \mathbb{S}_{mn}^{-1} \nonumber \\
  & & + \mathbb{S}_{ik}^{-1} \mathbb{S}_{jm}^{-1} \mathbb{S}_{ln}^{-1} 
      - \mathbb{S}_{ik}^{-1} \mathbb{S}_{jn}^{-1} \mathbb{S}_{lm}^{-1} 
      + \mathbb{S}_{il}^{-1} \mathbb{S}_{jk}^{-1} \mathbb{S}_{mn}^{-1}
      - \mathbb{S}_{il}^{-1} \mathbb{S}_{jm}^{-1} \mathbb{S}_{kn}^{-1}
      + \mathbb{S}_{il}^{-1} \mathbb{S}_{jn}^{-1} \mathbb{S}_{km}^{-1}
      - \mathbb{S}_{im}^{-1} \mathbb{S}_{jk}^{-1} \mathbb{S}_{ln}^{-1} \nonumber \\
  & & + \mathbb{S}_{im}^{-1} \mathbb{S}_{jl}^{-1} \mathbb{S}_{kn}^{-1}
      - \mathbb{S}_{im}^{-1} \mathbb{S}_{jn}^{-1} \mathbb{S}_{kl}^{-1} 
      + \mathbb{S}_{in}^{-1} \mathbb{S}_{jk}^{-1} \mathbb{S}_{lm}^{-1} 
      - \mathbb{S}_{in}^{-1} \mathbb{S}_{jl}^{-1} \mathbb{S}_{km}^{-1}
      + \mathbb{S}_{in}^{-1} \mathbb{S}_{jm}^{-1} \mathbb{S}_{kl}^{-1} \Big) \text{Pf} (\mathbb{S}) .
\end{eqnarray}
\end{subequations}
with the help of which we can obtain Eq. (\ref{I_3_final_b}) from Eq. (\ref{I_3_final_a}).
\end{widetext}



%

\end{document}